\documentclass[aps,prl,twocolumn,floatfix,longbibliography,superscriptaddress]{revtex4-2}

\usepackage{amsmath}
\usepackage{amssymb}
\usepackage{times}
\usepackage{braket}
\usepackage[pdftex]{graphicx}
\usepackage{color}
\usepackage{blindtext}
\usepackage{nicefrac}
\usepackage[colorlinks=true, citecolor=blue, urlcolor=blue, linkcolor=blue]{hyperref}

\usepackage{todonotes}

\renewcommand{\vec}[1]{\boldsymbol{#1}}

\newcommand{\change}[1]{\textcolor{black}{#1}}
\newcommand{\changeb}[1]{\textcolor{black}{#1}}

\renewcommand{\ket}[1]{\lvert#1\rangle}

\newcommand{\braopket}[3]{\langle #1 | #2 | #3\rangle}

\begin{document}

\title{Orbital Hall Effect Accompanying Quantum Hall Effect: \\Landau Levels Cause Orbital Polarized Edge Currents}

\author{B{\"o}rge G{\"o}bel}
\email[Correspondence email address: ]{boerge.goebel@physik.uni-halle.de}
\affiliation{Institut f\"ur Physik, Martin-Luther-Universit\"at Halle-Wittenberg, D-06099 Halle (Saale), Germany}

\author{Ingrid Mertig}
\affiliation{Institut f\"ur Physik, Martin-Luther-Universit\"at Halle-Wittenberg, D-06099 Halle (Saale), Germany}

\date{\today}

\begin{abstract}
The quantum Hall effect emerges when two-dimensional samples are subjected to strong magnetic fields at low temperatures: Topologically protected edge states cause a quantized Hall conductivity in multiples of $e^2/h$. Here we show that the quantum Hall effect is accompanied by an orbital Hall effect. Our quantum mechanical calculations fit well the semiclassical interpretation in terms of ``skipping orbits''. The chiral edge states of a quantum Hall system are orbital polarized akin to a hypothetical orbital version of the quantum anomalous Hall effect in magnetic systems. The orbital Hall resistivity scales quadratically with the magnetic field making it the dominant effect at high fields.
\end{abstract}

\maketitle


The quantum Hall effect is one of the most fundamental effects of modern solid-state physics. When a two-dimensional electron system is subjected to a strong magnetic field, the zero-field band structure turns into flat Landau levels~\cite{landau1930diamagnetismus,onsager1952interpretation,hofstadter1976energy}. In the 1970s, it was predicted and measured that these Landau levels give rise to a quantized Hall conductivity. 
von Klitzing \textit{et al.} determined these values to multiples of the constant $e^2/h$ and thereby revealed their fundamental significance~\cite{klitzing1980new}. 

The theoretical explanation of the quantum Hall effect is based on a non-trivial topology of the system~\cite{thouless1982quantized} induced by the magnetic field. The reciprocal-space Berry curvature~\cite{berry1984quantal,zak1989berry} of a perfectly flat Landau level is finite and the Landau levels can be characterized by an integer topological invariant~\cite{hatsugai2006topological,sheng2006quantum,gobel2017THEskyrmion,gobel2017QHE}. According to the bulk-boundary correspondence, topologically protected states emerge at the edge of the sample~\cite{hasan2010colloquium}. Each state contributes a conductivity quantum and bridges the gap between the Landau levels making the insulating sample conducting at the edge.

\begin{figure}[t!]
    \centering
    \includegraphics[width=\columnwidth]{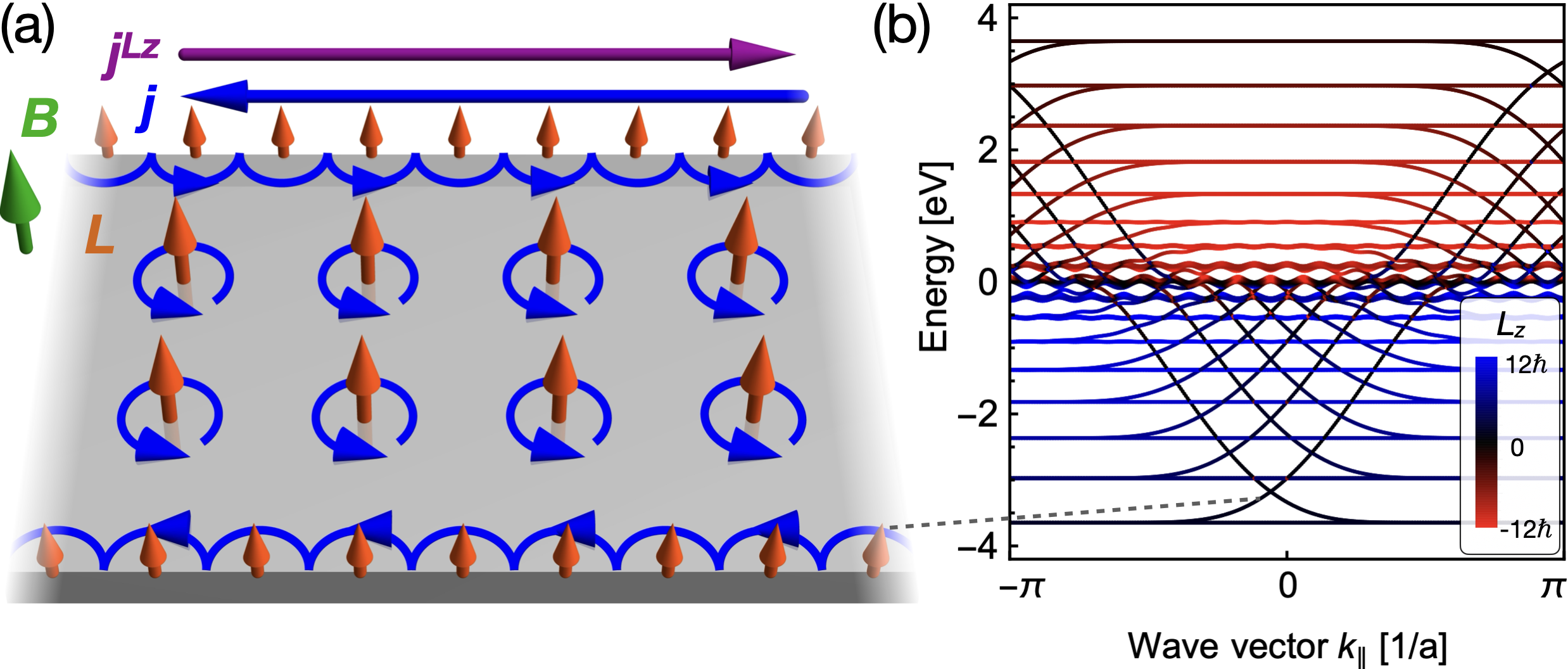}
    \caption{Orbital polarized edge currents in a quantum Hall system. (a) Blue represents electron trajectories affected by the magnetic field $\vec{B}$. Orbital angular momentum $\vec{L}$ is generated and skipping orbits emerge at the edges. They transport charge as charge currents $\vec{j}$ and orbital angular momentum as orbital currents $\vec{j}^{L_z}$ along the edges. (b) Band structure of a slab with $p/q=1/17$ and $q_0=6q=102$. Dispersive edge states are polarized with respect to the orbital angular momentum (color). They give rise to charge and orbital edge currents $\vec{j}$ and $\vec{j}^{L_z}$ \change{(dashed)}, as in (a). The left and right propagating states in one gap are located at opposite sides of the slab.}
    \label{fig:overview}
\end{figure}

In a semi-classical interpretation [cf. Fig.~\ref{fig:overview}(a)], moving electrons are forced onto circular trajectories in the bulk. However, at the edge, such a trajectory is not possible and `skipping orbits' emerge thereby transporting charge along the edge. The complete orbits in the bulk do not contribute to the conductivity but cause an out-of-plane orbital magnetization that can be calculated based on the modern formulation of orbital magnetization~\cite{chang1996berry,xiao2005berry,thonhauser2005orbital,ceresoli2006orbital,raoux2015orbital,gobel2018magnetoelectric}.

After the measurement and explanation of the quantum (charge) Hall effect, the quantum spin Hall effect was predicted and observed~\cite{kane2005quantum,kane2005z,bernevig2006quantum,konig2007quantum}. Instead of charge, spin angular momentum is transported along the edges. The combined spin and charge transport in magnetic materials is labeled anomalous Hall effect and it can also occur in a quantized manner~\cite{liu2008quantum,chang2013experimental,chang2015high}: The edge state transports charge and is spin polarized. More recently, interest has shifted also towards the orbital angular momentum~\cite{zhang2005intrinsic, bernevig2005orbitronics, kontani2008giant, tanaka2008intrinsic, kontani2009giant,go2018intrinsic, go2020orbital,go2021orbitronics,pezo2022orbital,canonico2020orbital,cysne2022orbital,salemi2022theory,busch2023orbital,choi2023observation,lyalin2023magneto,cysne2021disentangling,barbosa2023orbital,fonseca2023orbital,seifert2023time,busch2024ultrafast,johansson2021spin,el2023observation,go2020orbital,lee2021orbital}. While it is often quenched in equilibrium, it is considerable in transport phenomena. The orbital Hall effect\change{~\cite{zhang2005intrinsic, bernevig2005orbitronics, kontani2008giant, tanaka2008intrinsic, kontani2009giant,go2018intrinsic, go2020orbital,go2021orbitronics,pezo2022orbital,canonico2020orbital,cysne2022orbital,salemi2022theory,busch2023orbital,choi2023observation,lyalin2023magneto,cysne2021disentangling,barbosa2023orbital}} accompanies and often even surpasses the spin Hall effect and occurs even without spin-orbit coupling~\cite{go2018intrinsic, go2020orbital}. So far, the orbital Hall effect has mostly been discussed as a hybridization effect of different atomic orbitals~\cite{tanaka2008intrinsic, kontani2008giant, kontani2009giant, go2018intrinsic} but recently we have shown that $s$ orbitals can also generate an orbital Hall effect in a kagome lattice where the \change{effect stems from inter-site hybridization~\cite{busch2023orbital}. This contribution is crucial to describe the orbital transport in a quantum Hall system and it is included in the modern formulation of orbital magnetization~\cite{chang1996berry,xiao2005berry,thonhauser2005orbital,ceresoli2006orbital,raoux2015orbital,gobel2018magnetoelectric}. However, since it has been neglected by the community until recently, the quantum Hall effect and the skipping orbits have never been analyzed with respect to the orbital degree of freedom, so far.}

\begin{figure*}[t!]
    \centering
    \includegraphics[width=\textwidth]{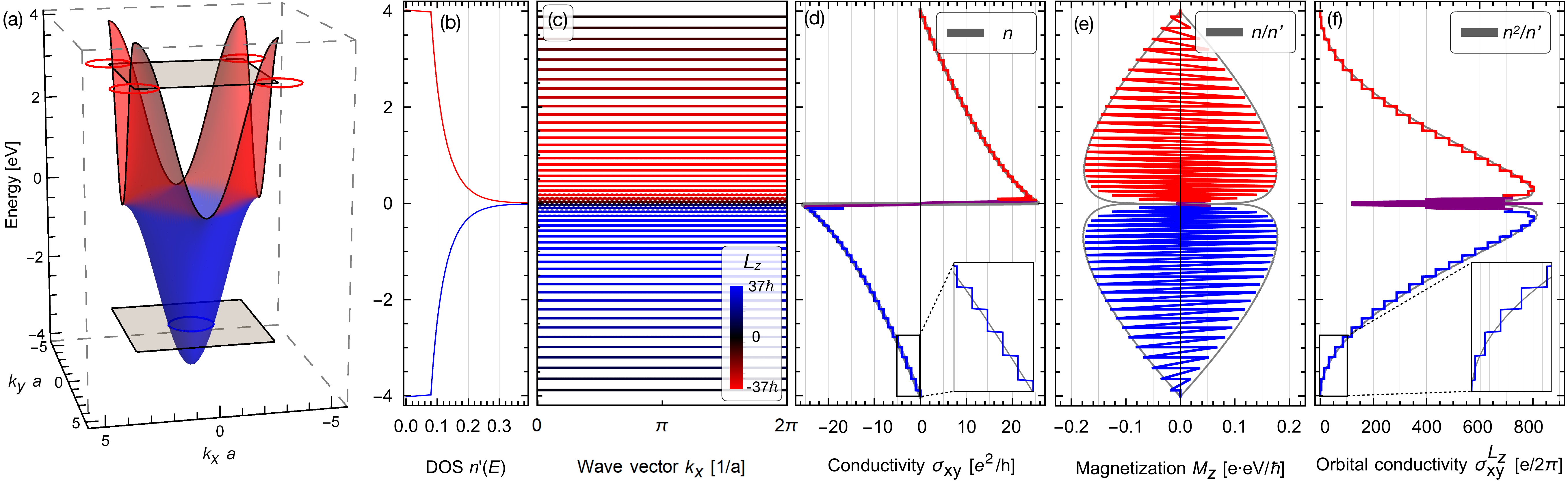}
    \caption{Orbital Hall effect accompanying the quantum Hall effect for $p/q=1/52$. (a) Zero-field band structure of the square lattice. The color indicates the effective mass of the charge carriers (blue: electron-like, red: hole-like). (b) Normalized density of states $n'(E)$ of the zero-field band structure. (c) Band structure under strong magnetic field $B\sim 80\,\mathrm{T}$ for $k_y=\pi/(2qa)$. Landau levels carry orbital angular momentum $L_z$ (blue: positive, red: negative). (d) Charge Hall conductivity. The gray curve resembles the scaled carrier density $n(E)$ for which the sign accounts for electron- and hole-like states. \change{(e) Orbital magnetization $M_z$. The gray curve resembles the scaled function $n(E)/n'(E)$. (f) Orbital Hall conductivity. The gray curve resembles the scaled function $[n(E)]^2/n'(E)$.}}
    \label{fig:calculations}
\end{figure*}

\begin{figure*}[t!]
    \centering
    \includegraphics[width=\textwidth]{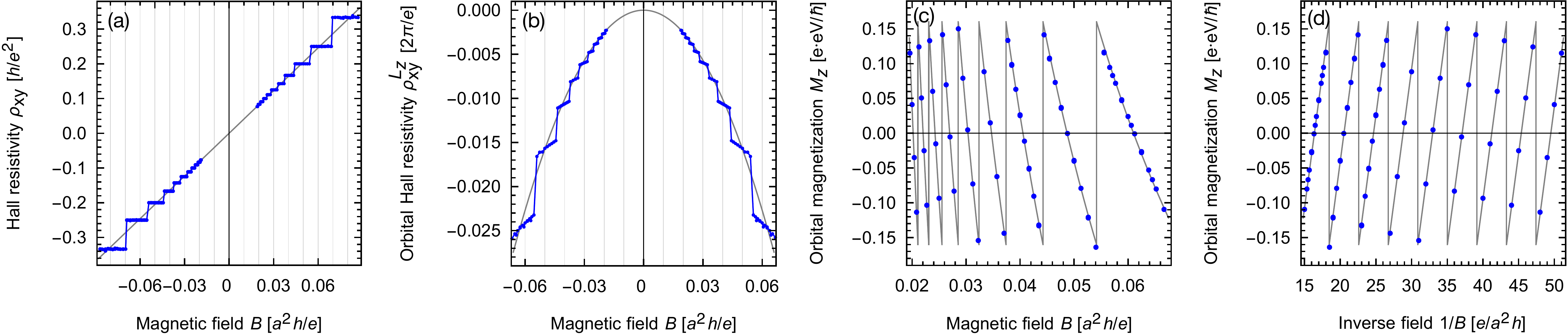}
    \caption{Magnetic field dependence. (a) Hall resistivity. The gray line shows the classical trend $\rho_{xy}\propto B_z$. (b) Orbital Hall resistivity. The gray line shows the trend $\rho_{xy}\propto -B_z^2$. (c) Orbital magnetization exhibiting De Haas–Van Alphen oscillations. The grey curves resemble the expected behavior for ideally flat Landau levels. (d) Inverse magnetic field dependence. Data points show results of our calculations for $q\leq 53$ and various $p\geq 1$ for a fixed $E_\mathrm{F}=-1.5\,\mathrm{eV}$.}
    \label{fig:observables}
\end{figure*}

In this Letter, we show \change{for the first time} that the quantum Hall effect is accompanied by an orbital Hall effect. Edge states bridge the gap between Landau levels in reciprocal space [cf. Fig.~\ref{fig:overview}(b)] and are polarized with respect to the orbital angular momentum. In real space, they transport charge and orbital angular momentum along the edge of the sample [cf. Fig.~\ref{fig:overview}(a)]. 
In the following, we use a tight-binding approach to calculate the intrinsic charge and orbital Hall conductivities as well as the orbital magnetization in the presence of a strong magnetic field based on the Berry theory~\cite{berry1984quantal} and the modern formulation of orbital magnetization~\cite{chang1996berry,xiao2005berry,thonhauser2005orbital,ceresoli2006orbital,raoux2015orbital,gobel2018magnetoelectric}. We discuss quantization effects and De Haas--van Alphen oscillations (Fig.~\ref{fig:calculations}) and determine the magnetic field dependence of the effects (Fig.~\ref{fig:observables}).
%


\textit{Landau levels in the presence of strong magnetic fields. --} We model a two-dimensional electron system under the effect of a strong magnetic field $\vec{B}=B\vec{e}_z$ by a tight-binding approach~\cite{hofstadter1976energy,claro1979magnetic, rammal1985landau, claro1981spectrum}. For clarity, we use a square lattice with one $s$ orbital per lattice site. 
%
%
Without the application of a magnetic field, the Hamiltonian directly reveals the band structure that consists of a single band [Fig.~\ref{fig:calculations}(a)].
Its density of states (DOS) $n'(E)$ is shown in Fig.~\ref{fig:calculations}(b). 

To simulate the magnetic field, we consider Landau gauge, with the vector potential $\vec{A} = -B y \vec{e}_{x}$ that enters the hopping terms of the tight-binding Hamiltonian. After a \change{Peierls} substitution, the hopping amplitude $t$ is transformed for each pair of neighboring lattice sites $(i,j)$ according to
\begin{align}
t \rightarrow t_0\exp\left(\frac{e}{\hbar}\int_{\vec{r}_i\rightarrow \vec{r}_j} \vec{A}(\vec{r})\cdot \mathrm{d}\vec{r}\right)
\end{align}
%
with $t_0=-1\,\mathrm{eV}$. 
%
These terms have to be periodic in order to be able to define a unit cell \change{of size $q_0$}. Therefore, coprime integers $p$ and $q$ have been introduced to define the magnetic field: $B=\frac{1}{a^2}\frac{h}{e}\frac{p}{q}$. 
Diagonalization of the Hamiltonian \change{(matrix form given in the Supplemental Material~\cite{SupplMat})} gives rise to the band structure $E_{\nu\vec{k}}$ and the eigenvectors $\ket{\nu\vec{k}}$.

The bulk band structure for $p/q=1/52$ is shown in Fig.~\ref{fig:calculations}(c). It consists of $q=52$ Landau levels. Even though the corresponding magnetic field of $B_z\sim 80\,\mathrm{T}$ (for \change{lattice constant} $a\sim 1\,\mathrm{nm}$) is very large, we have chosen this value because smaller fields would mean higher $q$ which would result in even more Landau levels which would make it impossible to resolve and properly discuss them. By contrast with free electrons, the emerging Landau levels are neither perfectly flat nor equidistant. This is because of the influence of the square lattice and the resulting zero-field band structure that is in general not parabolic [cf. Fig.~\ref{fig:calculations}(a)] and the DOS is not constant [cf. Fig.~\ref{fig:calculations}(b)]. 
Results for the cases $p/q=1/17$ and $p/q=3/52$ are shown in Supplemental Material, Figs. S1 and S2~\cite{SupplMat}.

Fig.~\ref{fig:overview}(b) shows the results of a slab calculation for \change{$p/q=1/17$} and $q_0=6q=102$. Again, we see $q=17$ rather flat bands corresponding to Landau levels that come in bundles of $5$ and form the bulk bands. However, since this time the geometry has actual edges, edge states emerge that bridge the band gaps. In the lowest band gap one right-propagating edge state  (mainly located at the upper edge) and one left-propagating state (at the lower edge) emerge . For each Landau level, one additional edge state emerges for each edge. This behavior corresponds to the edge channels giving rise to the quantum Hall effect. Since the effective mass of the zero-field band structure is positive for $E<0$, the charge carriers behave like electrons. Starting the discussion at the topmost band gap, the behavior is exactly opposite. This happens because the effective mass of the zero-field band structure is negative for $E>0$, so the charge carriers behave like holes.


\textit{Orbital angular momentum. --} Next, we will show that these edge charge currents are accompanied by orbital currents, so we introduce the $\vec{k}$-resolved orbital angular momentum along the out-of-plane direction. $L_{\nu,z}(\vec{k})$ are the diagonal elements of the tensor $\braopket{\nu \vec{k}}{L_z}{\alpha \vec{k}}$ where $\nu=\alpha$ is the band index. For further calculations of the transport properties, the off-diagonal tensor elements $\nu\neq\alpha$ are crucial as well. We calculate them based on the formalism derived by Pezo \textit{et al.}~\cite{pezo2022orbital} including a small correction to ensure gauge invariance of the orbital Hall conductivity
\begin{align}
      &\braopket{\nu \vec{k}}{L_z}{\alpha \vec{k}} = i  \frac{e\hbar^2}{4g_L\mu_\mathrm{B}}  \sum_{\beta \neq \nu, \alpha} \left( \frac{1}{E_{\beta \vec{k}} - E_{\nu \vec{k}}} + \frac{1}{E_{\beta \vec{k}} - E_{\alpha \vec{k}}} \right)\notag\\
      &\times\left(\braopket{\nu \vec{k}}{v_x}{\beta \vec{k}} \braopket{\beta \vec{k}}{v_y}{\alpha \vec{k}} - \braopket{\nu \vec{k}}{v_y}{\beta \vec{k}} \braopket{\beta \vec{k}}{v_x}{\alpha \vec{k}}\right) \label{EQ:Lz_matrix_elements}
\end{align}
Here, $\vec{v}=\frac{1}{\hbar}\nabla_{\vec{k}}H$ is the velocity operator. \change{The used approach is in accordance with the modern formulation of orbital magnetization~\cite{chang1996berry,xiao2005berry,thonhauser2005orbital,ceresoli2006orbital,raoux2015orbital,gobel2018magnetoelectric} that takes into account inter-site hybridization which is crucial for describing the quantum Hall system and can allow for finite $L_{\nu,z}(\vec{k})$ even for $s$ electrons that are characterized by $L_z=0$ in the atomic center approximation.}

The value of $L_{\nu,z}(\vec{k})$ has been used as a color code of the Landau levels in Fig.~\ref{fig:overview}(b) and Fig.~\ref{fig:calculations}(c). We see that for electron-like states for $E<0$, it is positive and for hole-like states for $E>0$, it is negative. In a semi-classical explanation, the negatively charged electrons are forced onto circular trajectories by the Lorentz force due to the magnetic field. For electron-like (hole-like) effective masses $m>0$ ($m<0$), the electrons move on a counter-clockwise (clockwise) trajectory giving rise to a positive (negative) orbital angular momentum along the out-of-plane direction $\vec{L}=\vec{r}\times\vec{p}$. Note that $\vec{L}$ of the edge states is smaller than in the bulk but still of considerable magnitude (multiple $\hbar$). At the edges, the orbits turn into skipping orbits giving rise to the edge channels that transport not only charge but also orbital angular momentum [cf. Fig.~\ref{fig:overview}(a)]. 


\textit{Observables: Orbital and quantum Hall conductivities and orbital magnetization. --} 
First we review the well known results of the quantum Hall conductivity $\sigma_{xy}(E_\text{F})$~\cite{nagaosa2010anomalous}
%
that is calculated from the reciprocal space Berry curvature $\Omega_{\nu,z}(\vec{k})$~\cite{berry1984quantal}
%
%
as a Brillouin zone integral over all occupied states \change{(see methods section~\cite{SupplMat})}. 
Fig.~\ref{fig:calculations}(d) shows the Hall conductivity that is quantized in units of $e^2/h$. \changeb{This is an intrinsic effect caused by the features of the tight-binding Hamiltonian.} Starting at the lowest discussed energy, it decreases step-wise to negative values upon increasing the energy. Upon crossing each Landau level, one conductivity quantum of $-e^2/h$ is added. This is because Landau levels have a Chern number of \change{$C_{\nu}=\int \Omega_{\nu,z}(\vec{k})\,\mathrm{d}^2k=1$} corresponding to the surface states discussed before in the slab band structure [Fig.~\ref{fig:overview}(b)]. At the energy of the van Hove singularity $E=0$ the Hall conductivity changes sign because the character of the charge carriers changes from electron-like to hole-like [cf. Fig.~\ref{fig:calculations}(a)]~\cite{lifshitz1957theory}. The Hall conductivity can be approximated by the carrier density $n(E)$
of the zero-field band structure, $\sigma_{xy,\mathrm{approx}}(E)\propto n(E)$ [cf. gray curve in Fig.~\ref{fig:calculations}(d)]~\cite{arai2009quantum,gobel2017QHE}.


While the magnetic field generates edge currents leading to a quantized Hall conductivity, it also generates a measurable orbital magnetization in the bulk that is calculated based on the modern formulation $M_z(E_\mathrm{F})$~\cite{chang1996berry,xiao2005berry,thonhauser2005orbital,ceresoli2006orbital,raoux2015orbital,gobel2018magnetoelectric}.
%
It is calculated as the integral over the diagonal elements of the orbital angular momentum $m_{\nu,z}(\vec{k})=-L_{\nu,z}(\vec{k})\cdot\frac{g_L\mu_B}{\hbar}$ plus a correction term that occurs due to the emergence of a reciprocal space Berry curvature~\cite{xiao2005berry}. 

Each Landau level contributes with its orbital angular momentum. However, the Berry curvature term becomes important especially in the band gaps where $\frac{\partial}{\partial E_\mathrm{F}}M_z(E_\mathrm{F})=-\frac{1}{2e}\sigma_{xy}(E_\mathrm{F})$. In \change{Fig.~\ref{fig:calculations}(e)} one can see that the two terms have opposite signs and lead to De Haas--Van Alphen oscillations
restricted by an envelope function~\cite{gat2003semiclassical}. We can determine it via $\Delta M_z = -\frac{1}{2e}\sigma_{xy}\Delta E$ and since $\sigma_{xy}\propto n$ and $\Delta E\propto 1/n'$, we find the envelope function of the orbital magnetization,
$M_{z,\mathrm{envelope}}(E_\mathrm{F})\propto \frac{n(E_\mathrm{F})}{n'(E_\mathrm{F})}$ (plotted as gray line).
The average orbital angular momentum in a Landau level $L_{\nu,z}$ can be approximated by the same energy dependence.

%


The orbital Hall conductivity~\cite{pezo2022orbital}
\begin{align}
    \sigma^{L_z}_{xy}(E_\text{F})= \frac{e}{\hbar}\sum_\nu \frac{1}{(2\pi)^2}\int_{E_{\nu \vec{k}}\leq E_\text{F}}\Omega_{\nu,z}^{L_z}(\vec{k}) \,\mathrm{d}^2k\label{EQ:sigma_Lz_xy_Kubo}
\end{align}
is calculated from the orbital Berry curvature~\change{\footnote{Note that the orbital Berry curvature results from linear response theory and is not an actual Berry curvature that gives rise to a topological invariant.}}
\begin{align}
    \Omega_{\nu,z}^{L_z}(\vec{k})= -2 \hbar^2\ \text{Im}\ \sum_{\mu\neq \nu} \frac{\braopket{\nu \vec{k}}{j_x^z}{\mu \vec{k}} \braopket{\mu \vec{k}}{v_y}{\nu \vec{k}}}{(E_{\nu \vec{k}} - E_{\mu \vec{k}})^2},
\end{align}
where $\braopket{\nu \vec{k}}{j_x^z}{\mu \vec{k}} = \frac{1}{2}\sum_{\alpha} [ \braopket{\nu \vec{k}}{v_x}{\alpha \vec{k}} \braopket{\alpha \vec{k}}{L_z}{\mu \vec{k}} + \braopket{\nu \vec{k}}{L_z}{\alpha \vec{k}} \braopket{\alpha \vec{k}}{v_x}{\mu \vec{k}} ]$
is the orbital current operator.

The orbital Hall conductivity is non-zero and exhibits several plateaus \change{[Fig.~\ref{fig:calculations}(f)]}. Starting from zero at the lowest discussed energy, it increases step-wise to positive values. Each Landau level leads to a step, however, with a different contribution for each Landau level. This means, the orbital Hall conductivity is constant in the gap between Landau levels but it is not quantized by some natural constant. At energies close to the van Hove singularity, several dispersive Landau levels are present that have a highly changing orbital Berry curvature. Consequently, the orbital Hall conductivity strongly changes with energy close to $E=0$. For positive energies beyond the van Hove singularity the orbital Hall conductivity decreases step-wise because the character of the charge carriers has changed from electron-like to hole-like.

The orbital Hall conductivity can be approximated by the product of the approximated Hall conductivity and the orbital magnetic moment
$\sigma_{xy,\mathrm{approx}}^{L_z}(E_{\mathrm{F}})\propto \frac{[n(E_\mathrm{F})]^2}{n'(E_\mathrm{F})}$.
To the best of our knowledge this has never been discussed before but even classically this is expected and corresponds to the conventional interpretation of the quantum Hall effect caused by skipping orbits [cf. Fig.~\ref{fig:overview}(a)]. A cycloid trajectory generates an out-of-plane orbital angular momentum and a charge current, so an orbital current emerges and is proportional to both. 
%
%

%
\change{\textit{Field dependence. --}}
Fig.~\ref{fig:observables}(a) shows the field dependence of the Hall resistivity $\rho_{xy}=-1/\sigma_{xy}$. Classically it exhibits a linear dependence on $B_z$; a behavior which is restored for large $q$ values. Here, for rather small fields, the Landau levels are so dense, that quantization effects become negligible. 
For smaller $q$ values, i.\,e., larger fields, the quantization becomes significant and the curve deviates from the linear dependence.

The orbital Hall resistivity $\rho_{xy}^{L_z}=-1/\sigma_{xy}^{L_z}$ exhibits a quadratic dependence on the magnetic field [cf. Fig.~\ref{fig:observables}(b)]: The larger $q$, the more Landau levels are occupied and the larger is their orbital angular momentum. Remarkably, the orbital Hall resistivity is negative independent of the sign of the magnetic field or the electron-like vs hole-like character of the charge carriers. The reason is that once $L_z$ changes, also the propagation direction of the edge currents changes and so the orbital edge current remains the same. 
For larger fields, quantization effects become significant and the curve deviates from the parabola. \change{Due to the continuous change of $L_z$ of the Landau levels with increasing magnetic field, $\rho_{xy}^{L_z}(B_z)$ does not exhibit plateaus in contrast to $\rho_{xy}^{L_z}(E_\mathrm{F})$. The orbital Hall resistivity changes with magnetic field even if the Fermi energy is still located in the same band gap.}

The orbital magnetization exhibits rather homogeneous De Haas--van Alphen oscillations [cf. Fig.~\ref{fig:observables}(c,d)] with $1/B\propto q/p$~\cite{farias2016haas}. Whenever a Landau level shifts through the Fermi energy, the orbital magnetization drops due to the orbital magnetic moment of a band $m_\nu^z(\vec{k})$, which is roughly constant with $\vec{k}$, especially when the Landau levels are flat.


\textit{Conclusion. --} Our calculations reveal that the edge states that are known to emerge under strong magnetic fields and exhibit the quantum Hall effect transport orbital angular momentum as well. The charge currents are accompanied by orbital currents and the quantum Hall effect is accompanied by an orbital Hall effect [cf. Fig.~\ref{fig:overview}(a)]. 

\change{The results persist even in a dirty sample with disorder. \changeb{Our calculations presented in Fig.~S3~\cite{SupplMat} show the vastly unchanged observables for one particular configuration of randomly varied on-site energies (no configuration averaging).} Furthermore, we have repeated the calculations on a triangular lattice and found similar results as for the square lattice (cf. Fig.~S4~\cite{SupplMat}).}

Our fundamental finding puts the characterization of the spintronic Hall effects~\cite{inoue2005taking,hirohata2020review} into a new perspective: One typically distinguishes (quantum) Hall, (quantum) spin Hall and (quantum) anomalous Hall effects characterized by the emergence of transverse charge currents, spin currents and spin-polarized charge currents, respectively. Orbital versions of the latter two effects have been discussed before~\cite{zhang2005intrinsic, bernevig2005orbitronics, kontani2008giant, tanaka2008intrinsic, kontani2009giant,go2018intrinsic, go2020orbital,go2021orbitronics,pezo2022orbital,canonico2020orbital,cysne2022orbital,salemi2022theory,busch2023orbital,choi2023observation,cysne2021disentangling,barbosa2023orbital}: Pure orbital currents can emerge in non-magnetic systems (orbital Hall effect) and orbital polarized charge currents in magnetic systems (orbital version of the anomalous Hall effect). Here, we have shown that the quantum Hall effect is actually not characterized by pure charge currents but by orbital polarized charge currents. In this regard, the quantum Hall effect is akin to the orbital version of the anomalous Hall effect even though the microscopic mechanisms are distinct. Both effects are characterized by the same reciprocal space $\vec{\Omega}_\nu$ and orbital Berry curvatures $\vec{\Omega}_\nu^{L_z}$ but the time-reversal symmetry $\mathcal{T}$ is broken by a magnetic field and a magnetic texture, respectively. 

To detect signatures of orbital currents experimentally, three approaches have been considered \changeb{(more details in~\cite{jo2024spintronics,SupplMat})}. (a) Orbital torques~\cite{lee2021orbital}: Orbital currents can be injected into an attached ferromagnet where the spin-orbital coupling generates a spin current that exerts a torque onto the ferromagnet that can be detected. (b) Magneto-optical Kerr effect (MOKE)~\cite{choi2023observation,lyalin2023magneto}: A Kerr rotation of reflected light can be measured to detect orbital angular momentum in particular at the edge of the sample and might reveal the orbital polarized edge states. (c) Inverse orbital effects~\cite{el2023observation}: Since the quantum Hall system exhibits an orbital Hall conductivity, an inverse effect exists as well. Orbital currents can be pumped from a ferromagnet and will be transformed into a charge current that can be measured electrically. To distinguish the orbital signature of the quantum Hall effect from the charge and spin contributions, it is convenient that the orbital Hall resistivity scales quadratically (neglecting the steps) with the magnetic field [cf. Fig.~\ref{fig:observables}(b)] by contrast to the roughly linear quantum Hall resistivity and the quantum anomalous Hall resistivity that switches only between two constant values ~\cite{he2015quantum,bestwick2015precise}. 



\section*{Acknowledgements}
This work was supported by the EIC Pathfinder OPEN grant 101129641 ``Orbital Engineering for Innovative Electronics''.

\bibliography{short,MyLibrary}

\end{document}